\documentclass{article}
\usepackage{arxiv}

\usepackage[utf8]{inputenc} 
\usepackage[T1]{fontenc}    
\usepackage{hyperref}       
\usepackage{url}            
\usepackage{booktabs}       
\usepackage{amsfonts}       
\usepackage{nicefrac}       
\usepackage{lipsum}		
\usepackage{graphicx}
\usepackage{natbib}
\usepackage{doi}
\usepackage{threeparttable}
\usepackage[nopatch]{microtype}
\usepackage{amsmath}
\usepackage{orcidlink}

\title{A Comparison of Joint and Stepwise Dynamic Cognitive Diagnostic Models}

\author{
  \href{https://orcid.org/0000-0001-7508-5385}{\includegraphics[scale=0.06]{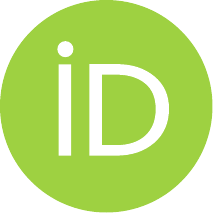}\hspace{1mm}Yawen Ma}\thanks{Corresponding author.} \\
 School of Mathematical Sciences \\
  Lancaster University \\
  Lancaster, LA1 4YF, United Kingdom \\
  \texttt{y.ma24@lancaster.ac.uk} \\
   \And
  \href{https://orcid.org/0000-0002-7930-6701}{\includegraphics[scale=0.06]{orcid.pdf}\hspace{1mm}Gabriel Wallin} \\
  School of Mathematical Sciences \\
  Lancaster University \\
  Lancaster, LA1 4YF, United Kingdom \\
  \texttt{g.wallin@lancaster.ac.uk} \\
  \And
  \href{https://orcid.org/0000-0002-0621-5032}{\includegraphics[scale=0.06]{orcid.pdf}\hspace{1mm}Anastasia Ushakova} \\
  Centre for Health Informatics, Computing, and Statistics \\
  Lancaster Medical School, Lancaster University \\
  Lancaster, LA1 4YW, United Kingdom \\
  \texttt{a.ushakova@lancaster.ac.uk} \\
  \And
  \href{https://orcid.org/0000-0003-2780-188X}{\includegraphics[scale=0.06]{orcid.pdf}\hspace{1mm}Kate Cain} \\
  Department of Psychology \\
  Lancaster University \\
  Lancaster, LA1 4YF, United Kingdom \\
  \texttt{k.cain@lancaster.ac.uk} \\
}

\renewcommand{\shorttitle}{Joint VS Stepwise Dynamic CDMs}

\hypersetup{
pdftitle={A template for the arxiv style},
pdfsubject={q-bio.NC, q-bio.QM},
pdfauthor={David S.~Hippocampus, Elias D.~Striatum},
pdfkeywords={First keyword, Second keyword, More},
}

\begin{document}
\maketitle

\begin{abstract}
	To extend cognitive diagnostic models (CDMs) to longitudinal settings, stepwise approaches that integrate a CDM model with a latent transition model and covariates are widely used due to their flexibility. Previous research has shown that stepwise estimation can yield biased results, motivating classification-error correction as a means of improving inference over uncorrected stepwise procedures. In this study, we evaluate a unified Bayesian dynamic cognitive diagnostic model that jointly estimates measurement (item parameters, latent attribute profiles) and transition components (transition parameters) in longitudinal settings with covariates. We compare this joint approach with the bias-corrected stepwise latent transition CDM through a Monte Carlo study. Results demonstrate that joint modeling provides more accurate recovery of transition parameters, particularly under limited test length and sample size, underscoring its advantages for longitudinal diagnostic analysis and offering practical guidance for applied researchers.
\end{abstract}


\section{Introduction}

Longitudinal cognitive diagnostic models (CDMs) provide a principled framework for examining changes in individuals' latent attribute profiles over time, focusing on discrete mastery transitions \citep{DeLaTorre2011}, rather than continuous latent growth trajectories (e.g., second-order growth model; \citep{Zheng2021}). By modeling change in dynamic latent variables rather than relying on cross-sectional representations of ability, longitudinal CDMs offer interpretable insights into learning processes, intervention effects, and  {changes in attribute profiles over time}. Consequently, these models have been increasingly applied in educational assessment \citep{Chen2017, Li2015, Wang2018a}, and psychological measurement \citep{Liang2024ThreeStepLTCDM}.

The evolution of longitudinal latent modeling has been strongly influenced by the development of stepwise estimation procedures, most notably the three-step approach. This framework typically involves: (1) estimating the latent measurement model (e.g., latent class analysis \citep{Goodman1974}, latent transition analysis \citep{Collins2010}, latent profile analysis \citep{Muthen2004}, or Gaussian Mixture Models \citep{Nylund2007, LiHarring2017} using only observed responses; (2) assigning individuals to latent classes while accounting for classification uncertainty; and (3) examining the associations between latent classes and external variables, such as covariates or distal outcomes \citep{Bakk2013, ClarkMuthen2009, Li2015}. Naive implementations of the three-step method treat assigned class memberships as error-free in step 2, leading to biased structural parameter estimates due to misclassification \citep{Croon2002}. A substantial body of methodological work addressed this limitation through bias-adjusted procedures designed to correct classification error using misclassification matrix or posterior probability \citep{Asparouhov2014, Bakk2013, Bakk2018, Bolck2004, Vermunt2010}. Because the classification error of the three-step method is introduced in its second step, alternative stepwise strategies have been proposed by eliminating this step; two-step approaches \citep{Bakk2018} retain separate estimation of the measurement model while avoiding explicit class assignment. Related developments trace back to earlier contributions \citep{Bartolucci2014, BandeenRoche1997, Xue2002}. In contrast, the one-step approach estimates the measurement and structural components simultaneously using full information maximum likelihood \citep{Clogg1981, Dayton1988, Lanza2013} or by incorporating logistic regression within DINA and higher-order DINA models to examine how covariates affect attribute mastery probabilities \citep{DelaTorre2004, Ayers2013, Park2014}. While statistically efficient, this one-step estimation can be computationally demanding, requires re-estimation of the entire model whenever structural components are modified, and blurs the distinction between measurement and structural parameters. These methodological advances highlight the importance of accounting for measurement uncertainty when examining relationships between latent classes and covariates. By estimating the measurement model independently, the latent class structure is protected from being altered by the inclusion of external variables. This separation further enables researchers to interpret the latent classes, particularly in educational contexts, to use socioeconomic and demographic variables, or other external covariates, to explain latent class membership. 

In many applications, researchers seek not only to describe change, but also to explain why learners differ in their initial mastery states and subsequent transitions. Individual characteristics, social demographics, and instructional interventions are therefore commonly incorporated as covariates to examine their effects on both initial attribute mastery and learning trajectories. This motivation has led to the development of longitudinal CDMs that integrate covariates into transition and structural components, most commonly through stepwise modeling strategies based on latent transition models \citep{Iaconangelo2016ThreeStepCDM, DiMari2016ThreeStepLM}. The integration of CDMs and latent transition models is attractive in practice because of its flexibility and ease of implementation. By contrast, one-step approaches estimate measurement and regression components jointly, but prior work has largely focused on cross-sectional settings \citep{Ayers2013, Park2014}, leaving limited evidence for their performance in longitudinal CDMs with covariates.

To address this gap, the present study compared a joint Bayesian dynamic CDM and a bias-corrected three-step latent transition CDM with covariates, under identical measurement specifications. Simulation studies examined differences in parameter recovery and attribute classification accuracy, providing practical guidance for applied longitudinal CDMs.

\section{Methodology Framework}
This study evaluates the joint dynamic CDM framework recently introduced in \citep{Ma2025} for modeling transitions between attribute profiles, and compares it with the bias-corrected three-step approach proposed by \citep{Liang2023ThreeStep}. Specifically, two estimation strategies were included in this study: 1) a joint Bayesian dynamic CDM that simultaneously estimates measurement and transition components, and 2) a bias-corrected three-step latent transition CDM with covariates. Both frameworks were examined using simulations under identical measurement specifications, including a common set of items administered at each time point, and time-invariant covariates. In the simulation study, the Q-matrix was assumed to be known and time-invariant for both estimation frameworks. To ensure interpretability and to focus the comparison on estimation strategy rather than measurement model choice, the deterministic-input, noisy-and-gate (DINA) model was adopted as the measurement model. In the following sections, we first briefly review DINA models and latent transition models, then introduce the bias-corrected stepwise modeling framework, and finally present the joint Bayesian dynamic formulation.

\subsection{A Brief Review of DINA Models}
CDMs belong to the family of latent class models (LCMs) and impose a specific structure on the latent space and on the relationship between latent classes and item responses. CDMs aim to characterize the relationship between learners’ latent attribute mastery profiles and their observed item responses. Let  {$\boldsymbol{\alpha}_{it} = (\alpha_{i1t}, \ldots, \alpha_{iKt})^\top$ denote the $K \times 1$ vector of binary attribute mastery profiles} for learner $i$ at time $t$ $(t = 1, \ldots,T)$, where $\alpha_{ikt} = 1$ indicates mastery of attribute $k $ $(k = 1,\ldots,K)$ and $\alpha_{ikt} = 0$ otherwise. The relationship between items and attributes is specified by the Q-matrix, where the element of Q-matrix $q_{jk} = 1$ indicates that the item $j$ requires attribute $k$.

Let $\mathbf{Y}_{it} = (Y_{i1t}, \ldots, Y_{iJt})^\top$
denote the $J \times 1$ vector of binary item responses for learner $i$ $(i=1,\ldots,N)$ at time $t$ $(t=1,\ldots,T)$, where $Y_{ijt}\in\{0,1\}$ indicates the response to item $j$ $(j=1,\ldots,J)$, with $1$ for a correct response and $0$ otherwise. In CDMs, item response probabilities are expressed through a latent indicator variable  {$\eta_{ijt}$}, which encodes whether learner $i$ meets the attribute requirements of item $j$  {at time $t$}. Conditional on  {$\eta_{ijt}$}, the probability of a correct response is given by
 
\begin{align}
\label{eq:dina}
P(Y_{ijt} = 1 \mid \boldsymbol{\alpha}_{it}) = (1 - s_j)^{\eta_{ijt}} g_j^{\smash{1 - \eta_{ijt}}},
\end{align}
where  {$s_j = P(Y_{ijt}=0 \mid \eta_{ijt}=1)$} and  {$g_j = P(Y_{ijt}=1 \mid \eta_{ijt}=0)$} are the slipping and guessing parameters for item $j$, respectively.

Under the deterministic-input, noisy-and-gate (DINA) model, {$\eta_{ijt}$} follows a conjunctive rule,  {reflecting the non-compensatory assumption that all required attributes must be mastered for a high probability of a correct response.}

\begin{equation*}
\eta_{ijt} = \prod_{k=1}^{K} \alpha_{ikt}^{q_{jk}},
\end{equation*}
so that  {$\eta_{ijt} = 1$} when learner $i$ has mastered all attributes required by item $j$  {at time $t$}.  

\subsection{Latent Transition Models with Covariates}
To model longitudinal change in attribute mastery, we adopt a latent transition framework. The latent attribute process is assumed to follow a first-order Markov structure, such that the attribute profile $\alpha_{ikt}$ at time $t$ depends on the profile at the previous time point $\alpha_{ik(t-1)}$. Covariates enter the model through the structural component, which affect both the initial attribute distribution and the transition probabilities. 
 {Let $\mathbf{Z}_{it} = (Z_{it,1},\ldots,Z_{it,C})^\top$ denote the $C \times 1$ vector of covariates for learner $i$ at time $t$.}
The initial mastery probability of attribute $k$ ($t = 1$) is modelled as
\begin{equation}
\label{eq:beta_initial}
\text{logit}\big(P(\alpha_{ik1}=1 \mid \mathbf{Z}_{i0})\big)
= \beta_{0k} + \sum_{c=1}^C \beta_{kc} Z_{i0,c},
\end{equation}
where $\beta_{0k}$ is the intercept and $\beta_{kc}$ represents the effect of covariate $\mathbf{Z}_{i0}$ on initial mastery of attribute $k$.

Attribute transitions from $t-1$ to $t$ are modelled conditionally on the previous attribute profiles and covariates. 
Specifically, the probabilities of acquiring and losing mastery for attribute $k$ are parameterized as
\begin{align}
\text{logit}\big(P(\alpha_{ikt}=1 \mid \alpha_{ik,t-1}=0, \mathbf{Z}_{i,t-1})\big)
&= \gamma_{01,k,0} + \sum_{c=1}^C \gamma_{01,k,c} Z_{i,t-1,c}, \label{eq:transition01} \\
\text{logit}\big(P(\alpha_{ikt}=0 \mid \alpha_{ik,t-1}=1, \mathbf{Z}_{i,t-1})\big)
&= \gamma_{10,k,0} + \sum_{c=1}^C \gamma_{10,k,c} Z_{i,t-1,c},
\label{eq:transition10}
\end{align}
where $\gamma_{01,k,c}$ denotes the effects of covariates $\mathbf{Z}_{i, t-1}$ on the probability of acquiring attribute $k$, and $\gamma_{10,k,c}$ denotes the effects of covariates on the probability of losing mastery of attribute $k$.

 {In the present study, we assume that once an attribute is mastered, it remains mastered over time, so mastery is treated as an absorbing state. This assumption is common in educational applications of longitudinal cognitive diagnosis, where learning is often viewed as cumulative and mastered attributes are unlikely to be unlearned within a short period \citep{Chen2017, YigitDouglas2021FirstOrder}. It is particularly reasonable in domains such as early reading development, where foundational skills (e.g., word decoding and vocabulary) serve as prerequisites for advanced skills such as comprehension \citep{Hoover}, and in spatial rotation tasks, where skill acquisition is typically progressive and stable over repeated practice \citep{Chen2017}. It is also a practically useful structural restriction because it reduces the number of possible learning trajectories and simplifies the transition states \citep{ChenCulpepper2020MVP}.  Importantly, both the stepwise and joint models can still be estimated without imposing the monotonicity assumption. Therefore, this assumption should not be interpreted as a requirement for identifiability; rather, it is a substantively motivated modeling choice that has been widely adopted in the literature \citep{LiuCulpepperChen2023}.}
This assumption reflects the cumulative nature of skill acquisition and helps distinguish learning progression from measurement error. Accordingly, attribute transitions are restricted to be monotonic in this study, such that, for each attribute $k$,
\begin{equation*}
P(\alpha_{ikt}=0 \mid \alpha_{ik,t-1}=1, \mathbf{Z}_{i,t-1}) = 0.
\end{equation*}

\subsection{Bias-Corrected Stepwise Latent Transition CDMs}
The bias-corrected stepwise approach separates the measurement and structural components and estimates them sequentially, while explicitly correcting for classification error induced by hard assignment of latent states. Following \citep{ DiMari2016ThreeStepLM, Liang2023ThreeStep}, the procedure consists of three steps, with measurement and transition specifications held identical to the joint framework presented in Section 2.4.

\subsubsection{Step 1: Measurement Model Estimation at Each Time Point}

At each time point, a DINA model is independently fitted to the response  {vector $\mathbf{Y}_{it}$ for learner $i$ at time $t$} under a pre-specified and time-invariant Q-matrix. Fitting the DINA model provides estimates of the item parameters and the posterior distribution over attribute profiles for each learner,
\begin{equation*}
 {
\widehat{p}_{it}(\boldsymbol{\alpha})
=
P(\boldsymbol{\alpha}_{it}=\boldsymbol{\alpha} \mid \mathbf{Y}_{it}, \widehat{\mathbf{s}}, \widehat{\mathbf{g}}, \mathbf{Q}),
\qquad 
\boldsymbol{\alpha}\in\{0,1\}^K.
}
\end{equation*}

\subsubsection{Step 2: Latent State Assignment and Classification Error Estimation}

To facilitate subsequent modeling of transitions and covariate effects, learners are assigned to discrete attribute  {profiles} based on the posterior distributions obtained in Step 1. Specifically, a maximum a posteriori (MAP) assignment rule is applied such that the estimated attribute profile for learner $i$ at time $t$ is defined as 
\begin{equation*}
 {
\widehat{\boldsymbol{\alpha}}_{it}
=
\arg\max_{\boldsymbol{\alpha}\in\{0,1\}^K}
\widehat{p}_{it}(\boldsymbol{\alpha}).
}
\end{equation*}
The resulting assigned $\widehat{\boldsymbol{\alpha}}_{it}$ are treated as observed indicators of attribute profiles and are used as inputs to the structural model. However, this hard assignment ignores posterior uncertainty and introduces misclassification error. To account for this error, classification error probabilities (CEPs) are introduced.  {Let $\mathbf{W}_i=(W_{i1},\ldots,W_{iT})^\top$ denote the assigned attribute
profiles across time.} The CEP matrix ($\mathbf{M}_t$) summarizes the misclassification induced by the stepwise assignment, linking the unobserved true attribute  {profiles} $L_{it}$ to the observed assigned attribute  {profiles} $W_{it}$ at each time point $t$. The $\mathbf{M}_t$ is defined as
\begin{equation*}
\mathbf{M}_t(r,s)
=
P(W_{it}=r \mid L_{it}=s),
\qquad r,s \in \mathcal{S}, \mathcal{S}=\{0,1\}^K,
\end{equation*}
where $\mathcal{S}$ denotes the set of all possible attribute profiles, with $|\mathcal{S}|=2^K$. For example, when $K=2$, $\mathcal{S}=\{(0,0),(0,1),(1,0),(1,1)\}$. Here, $L_{it}=s$ indicates that the value of the true attribute vector
$\boldsymbol{\alpha}_{it}$ for learner $i$ at time $t$ is $s$, while $W_{it}=r$ denotes the attribute profile $r$ assigned by Step 1. Hence, the entry $M_t(r,s)$
represents the probability that a learner with true attribute profile $s$
is assigned to profile $r$ at Step 1. In practice, $\mathbf{M}_t$ is estimated by averaging posterior classification probabilities obtained in Step 1 over individuals,
\begin{equation*}
\widehat{\mathbf{M}}_t(r,s)
=
\frac{
\sum_{i=1}^{N}
P(L_{it}=s \mid \mathbf{Y}_{it}) \,
\mathbf{I}(W_{it}=r)
}{
\sum_{i=1}^{N}
P(L_{it}=s \mid \mathbf{Y}_{it})
}.
\end{equation*}

\subsubsection{Step 3: Structural Model Estimation with CEP Correction}
In the final step, the latent transition model with covariates is estimated while correcting for misclassification in attribute  {profiles}. The structural model specifies the initial attribute distribution  {$P(L_{i1} \mid \mathbf{Z}_{i0})$} and the transition probabilities  {$P(L_{it} \mid L_{i,t-1}, \mathbf{Z}_{i,t-1})$}, allowing covariates to influence both initial mastery and transitions. Because the true latent states $L_{it}$ are unobserved, estimation is carried out by marginalizing over all possible latent state sequences while conditioning on the observed assignments $W_{it}$. 
The likelihood contribution for individual $i$ is given by
\begin{equation}
P(\mathbf{W}_i \mid \mathbf{Z}_i)
=
\sum_{L_{i1}}\cdots\sum_{L_{iT}}
P(L_{i1}\mid \mathbf{Z}_{i0})
\prod_{t=2}^T P(L_{it}\mid L_{i,t-1}, \mathbf{Z}_{i,t-1})
\prod_{t=1}^T \mathbf{M}_t(W_{it}, L_{it}).
\label{eq:three_step_lik}
\end{equation}
This formulation explicitly incorporates classification error by weighting
each possible attribute profile according to the CEP matrix estimated in
Step 2. The structural parameters (initial mastery probability and transition probabilities) are estimated by maximizing the likelihood
in Equation \eqref{eq:three_step_lik}.

\subsection{A Joint Bayesian Dynamic CDM}
In contrast, the joint approach retains the same separation between measurement and structural components at the model specification level, but jointly estimates all components within a joint Bayesian estimation framework. Specifically, latent attribute profiles ($\boldsymbol{\alpha}_{it}$), item parameters ($g_j$, $s_j$), initial attribute mastery parameters ($\beta_{0k}$, $\beta_{kc}$), and transition parameters ($\gamma_{01,k,0}$, $\gamma_{01,k,c}$) are estimated simultaneously using Markov  {chain} Monte Carlo (MCMC) sampling.

\subsubsection{Prior Specifications}
 {Weakly informative priors are assigned to all model parameters to ensure that posterior inference is primarily driven by the likelihood while allowing sufficient flexibility. For the guessing and slipping parameters, we used non-informative priors (flat priors) following \citep{wang2018b}. Specifically, for item parameters, independent priors} were assigned to the guessing and slipping parameters,
\begin{align*}
g_j \sim \text{Beta}(1, 1), \qquad 
s_j \sim \text{Beta}(1, 1).
\end{align*}
Regression coefficients for initial attribute mastery and transition processes were assigned  {normal priors $N(0,1)$:}
\begin{align*}
 {\beta_{0k},\ \beta_{kc} \sim \mathcal{N}(0, 1),} \qquad
 {\boldsymbol{\gamma}_{01,k,0},\boldsymbol{\gamma}_{01,k,c} \sim \mathcal{N}(0, 1).}
\end{align*}
 {All continuous covariates were standardized to have mean zero and unit variance; under this scaling, these priors are weakly informative. For the transition parameters $\boldsymbol{\gamma}_{10,k,0}$ and $\boldsymbol{\gamma}_{10,k,c}$, we specified more informative priors for the intercepts centered at $-2$ on the logit scale to reflect the absorbing learning assumption, while retaining weakly informative $\mathcal{N}(0,1)$ priors for the covariate effects following \citep{Ma2025}.}
Latent attribute profiles are treated as random variables and inferred jointly with all other parameters.

\subsubsection{Inference}

Inference is conducted by sampling from the joint posterior distribution of all unknown quantities given the observed data. Specifically, the joint posterior distribution is given by
\begin{align*}
P(\mathbf{g}, \mathbf{s}, \boldsymbol{\beta}, \gamma_{01}, \gamma_{10},
\boldsymbol{\alpha}_1, \ldots, \boldsymbol{\alpha}_T
\mid \mathbf{Y}_1, \ldots, \mathbf{Y}_T, \mathbf{Z}, \mathbf{Q})
\propto
& \prod_{t=1}^T P(\mathbf{Y}_t \mid \mathbf{Q}, \mathbf{g}, \mathbf{s}, \boldsymbol{\alpha}_t) \notag \cdot  \\
& \prod_{t=2}^T
P(\boldsymbol{\alpha}_t \mid \boldsymbol{\alpha}_{t-1}, \gamma_{01}, \gamma_{10}, \mathbf{Z}_{t-1}) \notag \cdot\\
& P(\boldsymbol{\alpha}_1 \mid \boldsymbol{\beta}, \mathbf{Z}_0) \cdot\,
P(\mathbf{g}, \mathbf{s}) \cdot\,
P(\boldsymbol{\beta}, \gamma_{01}, \gamma_{10}),
\end{align*}
 {where $\mathbf{Y}_t = \{Y_{ijt}\}_{i=1,\ldots,N;\, j=1,\ldots,J}$, and $\boldsymbol{\alpha}_t = (\boldsymbol{\alpha}_{1t},\ldots,\boldsymbol{\alpha}_{Nt})$.} 
The first term,  {$P(\mathbf{Y}_t \mid \mathbf{Q}, \mathbf{g}, \mathbf{s}, \boldsymbol{\alpha}_t)$}, represents the likelihood of observed responses at time $t$ based on DINA model defined in Equation \eqref{eq:dina}. The second term, $P(\boldsymbol{\alpha}_t \mid \boldsymbol{\alpha}_{t-1}, \gamma_{01}, \gamma_{10}, \mathbf{Z}_{t-1})$, specifies the transition probabilities between attribute profiles across time {, characterized by the acquisition and loss mastery models} as defined in \eqref{eq:transition01} and \eqref{eq:transition10}. While the third term, $P(\boldsymbol{\alpha}_1 \mid \boldsymbol{\beta}, \mathbf{Z}_0)$, models the initial attribute mastery distribution in Equation \eqref{eq:beta_initial}. The remaining terms $P(\mathbf{g}, \mathbf{s})$ and $P(\boldsymbol{\beta}, \gamma_{01}, \gamma_{10})$, represent prior distributions on item parameters and regression coefficients, respectively. 

\subsubsection{Computation}
For the stepwise framework, we followed a three-step latent transition cognitive diagnostic modeling strategy \citep{Liang2023ThreeStep}. Specifically, DINA measurement models were fitted separately at each time point using the R package GDINA \citep{MaDeLaTorre2020GDINA}. The CEPs were then estimated and incorporated to correct for misclassification. Finally, covariate effects on initial attribute mastery and attribute transitions were estimated via logistic regression, corresponding to Equations \eqref{eq:beta_initial}, \eqref{eq:transition01}, and \eqref{eq:transition10}. 

For joint modeling, posterior inference was conducted using MCMC sampling implemented in JAGS (Just Another Gibbs Sampler) via the R package RJags. Full conditional distributions were derived for all parameters, and samples were drawn iteratively to approximate the joint posterior distribution. Convergence was assessed using the potential scale reduction factor ($\widehat{R}$) and visual inspection of trace plots. The DINA implementation follows standard cognitive diagnostic modeling practice, see \citep{Zhan2019JEBStats} for further examples of RJAGS-based CDMs. The code used for the simulation study is publicly available, see \ref{sec:app_code} for details.

\section{Simulation}

\subsection{Simulation Design}

We conducted a simulation study to compare joint and stepwise estimation frameworks for longitudinal CDMs with time-invariant covariates. The design included two time points and two latent attributes, with item parameters held constant across time to ensure longitudinal measurement invariance. The Q-matrix was assumed to be known and time-invariant for both estimation frameworks. Three combinations of sample size and test length were considered, (N, J) = (200, 6), (400, 18), (600, 30), and the joint and stepwise estimation approaches were compared across these conditions. Regarding the sample size and test length conditions, these values were selected to represent small to moderate testing scenarios commonly encountered in cognitive diagnostic assessments \citep{culpepper2015,delatorre2009}. Moreover, the same design settings were adopted as in our previous study under the Q-matrix unknown framework \citep{Ma2025}, allowing for a direct and meaningful comparison between the current results and prior findings.

For each learner $i$, a vector of three time-constant covariates $\mathbf{Z}_i$ was generated and incorporated into both the initial attribute mastery and transition models. To better reflect realistic dependence among predictors, these covariates were generated from a correlated distribution, such that the covariate vector $\mathbf{Z}_i = (Z_{i1}, Z_{i2}, Z_{i3})^\top$ follows a three-dimensional multivariate normal (MVN) distribution,
\begin{align*}
\mathbf{Z}_i \sim \text{MVN}(\mathbf{0}, \Sigma),
\end{align*}
where
$$
\Sigma =
\begin{pmatrix}
1 & \rho & \rho \\
\rho & 1 & \rho \\
\rho & \rho & 1
\end{pmatrix}.
$$

In the main simulation results, we set $\rho = 0.4$ to represent a moderate level of correlation among covariates. Sensitivity analyses for the most challenging simulation condition $(N, J) = (200, 6)$ under varying correlation levels $\rho \in \{0, 0.2, 0.4, 0.6, 0.8\}$ are provided in \ref{sec:sensitivity_cov} (Tables \ref{tab:cov} and \ref{tab:cov1}).
Item responses at each time point were generated under the DINA model using a fixed Q-matrix that satisfies the necessary and sufficient conditions for model identifiability \citep{GuXu2021Identifiability}. The Q-matrices used in simulations are provided in the \ref{sec:app_Qmatrix}. Guessing and slipping parameters were held constant across time. According to \citep{MaIaconanglo2016}, item quality was set at a moderate level by independently drawing guessing and slipping parameters from a $\text{Uniform}(0.15, 0.25)$ distribution.

Model performance was evaluated in terms of parameter recovery and classification accuracy. For regression and transition parameters, mean absolute error (MAE) and root mean square error (RMSE) were computed across replications, while classification accuracy was assessed using the attribute-wise agreement rate (AAR) at each time point, a standard metric in cognitive diagnostic modeling (e.g., \citealp{ChiuDouglas2013ProximityCDM, ZhanJiaoLiao2018CDMRT}). Definitions of the MAE and RMSE metrics are provided in the \ref{sec:metric}. 
Specifically, AAR for attribute $k$ at time point $t$ was defined as
$$
\text{AAR}_{k,t} = \frac{1}{NR} \sum_{r=1}^{R} \sum_{i=1}^{N} \mathbb{I} \bigl(\hat{\alpha}^{(r)}_{ikt} = \alpha_{ikt}\bigr),
$$
where $\alpha_{ikt}\in\{0,1\}$ is the true mastery state and $\hat{\alpha}^{(r)}_{ikt}$ is its estimate in replication $r$.

\subsection{Simulation Results}
We conducted 100 replications per simulation condition. We fitted the joint Bayesian model in JAGS using two parallel chains. After 1,000 burn-in iterations, we drew 2,000 MCMC iterations per chain. Convergence was assessed using the potential scale reduction factor ($\hat{R}$; \citep{gelman1992inference}), and the maximum $\hat{R}$ across key parameters was below 1.2, with approximately 98\% of monitored parameters having PSRF values less than 1.02. The Monte Carlo error (MCE) for coverage probability is 0.02 across all conditions following \citep{KoehlerBrownHaneuse2009}, confirming 100 replications are sufficient. Using a single CPU core, runtimes across conditions for each iteration ranged from approximately 0.8 to 6 minutes. Parameter estimates were obtained by averaging the estimates from two parallel chains with randomly chosen starting values for each replication, and summarizing over 100 replications. 

Table \ref{tab:alpha_aar} reports attribute recovery accuracy (AAR) for the joint and stepwise approaches across simulation conditions. Both methods achieved high AARs, with recovery accuracy increasing as sample size and test length increased. The joint approach showed slightly higher AARs under the smallest sample condition ($N=200, J=6$). For the large sample condition, both approaches achieved high recovery accuracy, with AARs exceeding 0.99 across attributes and time points. Table \ref{tab:gs_bias} in \ref{sec:app_item} summarises the estimation accuracy of guessing and slipping parameters ($g_{jt}$ and $s_{jt}$) under the joint and stepwise frameworks. Estimation accuracy improved with increasing sample size and test length, with RMSEs for $g_{jt}$ and $s_{jt}$ substantially lower under moderate and large conditions. Across all conditions, MAE values remained small, generally below 0.06 for the joint approach and below 0.07 for the stepwise approach, indicating accurate recovery of item parameters for both methods.

\begin{table}[!h]
\centering
\caption{\label{tab:alpha_aar}Attribute recovery (AAR) under Joint and Stepwise.}
\begin{threeparttable} 
\begin{tabular}{ccc|cc|cc}
\toprule
\multicolumn{3}{c}{} & \multicolumn{2}{c}{Joint} & \multicolumn{2}{c}{Stepwise} \\
\cmidrule(l{3pt}r{3pt}){4-5} \cmidrule(l{3pt}r{3pt}){6-7}
$N$ & $J$ & $T$ & AAR1 & AAR2 & AAR1 & AAR2 \\
\midrule
200 & 6  & 1 & 0.899 & 0.869 & 0.868 & 0.846 \\
    &    & 2 & 0.891 & 0.858 & 0.861 & 0.843 \\
400 & 18 & 1 & 0.978 & 0.972 & 0.971 & 0.973 \\
    &    & 2 & 0.974 & 0.968 & 0.970 & 0.969 \\
600 & 30 & 1 & 0.993 & 0.992 & 0.993 & 0.992 \\
    &    & 2 & 0.994 & 0.993 & 0.992 & 0.992 \\
\bottomrule
\end{tabular}
\begin{tablenotes}[flushleft]
\footnotesize
\item[] \textit{Note.} AAR1 and AAR2 denote attribute-wise recovery accuracy for Attribute 1 and Attribute 2, respectively.
\end{tablenotes}
\end{threeparttable} 
\end{table}

Table \ref{tab:beta_metrics} summarizes the estimation accuracy of initial covariate effects ($\beta$) under the joint and stepwise frameworks. Across all conditions, both MAE and RMSE decreased as sample size and test length increased. For the smallest condition ($N=200$, $J=6$), notable estimation error was observed particularly for intercept terms, whereas slope estimates were more stable. Both approaches exhibited similar accuracy, although the joint framework tended to yield slightly smaller errors for parameters under smaller sample sizes.

\begin{table}[!h]
\centering
\caption{\label{tab:beta_metrics}Estimation accuracy of initial covariate effects ($\beta$) evaluated by MAE and RMSE across conditions.}
\begin{threeparttable} 
\begin{tabular}{ccc|cccc|cccc}
\toprule
\multicolumn{3}{c}{} & \multicolumn{4}{c}{Joint} & \multicolumn{4}{c}{Stepwise} \\
\cmidrule(l{3pt}r{3pt}){4-7}\cmidrule(l{3pt}r{3pt}){8-11}
$N$ & $J$ & Metric 
& $\beta_{0,1}$ & $\beta_{0,2}$ & $\beta_{Z,1}$ & $\beta_{Z,2}$ 
& $\beta_{0,1}$ & $\beta_{0,2}$ & $\beta_{Z,1}$ & $\beta_{Z,2}$ \\
\midrule
200 & 6  & MAE  
& 0.182 & 0.366 & 0.055 & 0.095 
& 0.229 & 0.279 & 0.060 & 0.105 \\
    &    & RMSE 
& 0.197 & 0.225 & 0.067 & 0.105 
& 0.275 & 0.317 & 0.067 & 0.119 \\
\addlinespace
400 & 18 & MAE  
& 0.094 & 0.085 & 0.055 & 0.042 
& 0.099 & 0.093 & 0.071 & 0.055 \\
    &    & RMSE 
& 0.108 & 0.106 & 0.065 & 0.051 
& 0.120 & 0.126 & 0.085 & 0.068 \\
\addlinespace
600 & 30 & MAE  
& 0.060 & 0.079 & 0.031 & 0.023 
& 0.057 & 0.073 & 0.024 & 0.031 \\
    &    & RMSE 
& 0.071 & 0.092 & 0.039 & 0.028 
& 0.082 & 0.094 & 0.034 & 0.040 \\
\bottomrule
\end{tabular}
\begin{tablenotes}[flushleft]
\footnotesize
\item \textit{Note.} $\beta_{0,k}$ denotes intercepts and $\beta_{Z,k}$ denotes covariate effects for Attribute $k$.
\end{tablenotes}
\end{threeparttable} 
\end{table}

Table \ref{tab:gamma_slope_metrics} summarises the estimation accuracy of transition-related covariate effects on attribute acquisition ($\gamma_{01,Z,k}$) under the joint and stepwise frameworks. Across all conditions, the joint model consistently achieved lower MAE and RMSE than the stepwise approach, with differences being observed under small sample sizes ($N=200$) and short tests ($J=6$). Estimation accuracy improved as $N$ and $J$ increased for both methods; however stepwise estimates of $\gamma_{01}$ remained considerably more variable even in larger samples. Sensitivity analyses reported in \ref{sec:sensitivity_cov} further suggest that the joint approach shows limited sensitivity to varying covariate correlation levels $\rho$, with MAE and RMSE for item parameters and regression parameters typically changing by less than 0.1 relative to the independent covariate setting ($\rho = 0$). Because mastery is treated as an absorbing state through an informative prior on $\gamma_{10}$ that makes attribute loss effectively impossible, these parameters are not of substantive interest and are not reported.

\begin{table}[!h]
\centering
\caption{Estimation accuracy of transition covariate effects ($\gamma_{01}$ slopes) evaluated by MAE and RMSE across conditions.}
\label{tab:gamma_slope_metrics}
\begin{threeparttable} 
 {
\begin{tabular}{ccccccc}
\toprule
\multicolumn{3}{c}{} &
\multicolumn{2}{c}{Joint} &
\multicolumn{2}{c}{Stepwise} \\
\cmidrule(l{3pt}r{3pt}){4-5}
\cmidrule(l{3pt}r{3pt}){6-7}
$N$ & $J$ & Metric &
$\gamma_{01,Z,1}$ & $\gamma_{01,Z,2}$ &
$\gamma_{01,Z,1}$ & $\gamma_{01,Z,2}$ \\
\midrule
200 & 6 & MAE
& 0.100 & 0.140
& 1.008 & 0.805 \\
    &   & RMSE
& 0.125 & 0.164
& 1.208 & 0.911 \\
\addlinespace
400 & 18 & MAE
& 0.060 & 0.064
& 0.533 & 0.489 \\
    &    & RMSE
& 0.085 & 0.077
& 0.629 & 0.570 \\
\addlinespace
600 & 30 & MAE
& 0.063 & 0.058
& 0.489 & 0.381 \\
    &    & RMSE
& 0.073 & 0.070
& 0.674 & 0.457 \\
\bottomrule
\end{tabular}}
\begin{tablenotes}[flushleft]
\footnotesize
\item \textit{Note.} $\gamma_{01,Z,k}$ denotes the covariate effect $Z$ on the transition probability from non-mastery to mastery for Attribute $k$. 
\end{tablenotes}
\end{threeparttable} 
\end{table}

\section{Discussion \& Future Directions}
Across simulation conditions, the joint framework demonstrated better overall estimation performance relative to the stepwise framework, particularly in the recovery of covariate effects on attribute acquisition. This advantage was most evident under small sample and short test conditions, where joint modeling provided more stable and accurate estimates than stepwise procedures. Previous research has noted that transition parameters are generally more difficult to estimate accurately than initial state parameters under the stepwise framework \citep{Liang2023ThreeStep}. This pattern likely reflects the cumulative impact of classification uncertainty in stepwise procedures, where transition models are built on latent state assignments obtained from earlier stages. Findings of this study indicated that the proposed joint model improved the accuracy of transition parameter recovery. Notably, the joint framework achieved these gains while remaining computationally efficient, supporting its practical applicability in longitudinal diagnostic settings.

As noted by \citep{Liang2023ThreeStep}, a key advantage of stepwise approaches is their flexibility at the measurement stage. Response data are modelled separately at each time point, allowing measurement model specification and model selection to be conducted independently. Future work should investigate extensions of the joint framework proposed in this paper to alternative measurement models, such as deterministic input noisy "or" gate \citep{templin2006measurement}, Generalized DINA \citep{de2016general}, and higher-order DINA \citep{dela2004}. Additionally, future work could consider the inclusion of time-varying covariates. 

From a broader methodological perspective, most existing longitudinal CDMs assume a known or fixed Q-matrix. In practice, however, the Q-matrix is often unknown, and researchers have developed various methods to estimate the Q-matrix from response data (see \citep{ChenCulpepper2017DINAQ, Chung2019GibbsQ, Wang2020}). For example, \citep{Ma2025} investigate how data-driven Q-matrix estimation can be integrated within unified longitudinal CDM frameworks. Extending such joint approaches to real-world educational data represents an important direction for applied research, especially for large-scale assessments such as the Programme for International Student Assessment (PISA), the Programme for the International Assessment of Adult Competencies (PIAAC), and the Trends in International Mathematics and Science Study (TIMSS).

\paragraph{Funding Statement}

This research was supported by the Engineering and Physical Research Council, which funded the first author through a PhD studentship. 

\paragraph{Competing Interests}

None.


\bibliographystyle{unsrt}
\bibliography{references.bib}  

@article{delatorre2009,
  author  = {de la Torre, Jimmy},
  title   = {DINA Model and Parameter Estimation: A Didactic},
  journal = {Journal of Educational and Behavioral Statistics},
  year    = {2009},
  volume  = {34},
  number  = {1},
  pages   = {115--130},
  doi     = {10.3102/1076998607309474}
}

@article{culpepper2015,
  author  = {Culpepper, Steven A.},
  title   = {Bayesian Estimation of the DINA Model},
  journal = {Journal of Educational and Behavioral Statistics},
  year    = {2015},
  doi     = {10.3102/1076998615595403}
}

@ARTICLE{KoehlerBrownHaneuse2009,
  author = {{Koehler}, Elizabeth and {Brown}, Elizabeth and {Haneuse}, Sebastien J.-P. A.},
  title = {On the Assessment of Monte Carlo Error in Simulation-Based Statistical Analyses},
  journal = {The American Statistician},
  year = 2009,
  volume = 63,
  number = 2,
  pages = {155--162},
  doi = {10.1198/tast.2009.0030}
}

@ARTICLE{LiuCulpepperChen2023,
  author = {{Liu}, Yang and {Culpepper}, Steven A. and {Chen}, Yinyin},
  title = {Identifiability of Hidden Markov Models for Learning Trajectories in Cognitive Diagnosis},
  journal = {Psychometrika},
  year = 2023,
  volume = 88,
  number = 2,
  pages = {361--386},
  doi = {10.1007/s11336-023-09904-x}
}

@article{Hoover,
author = {Hoover, Wesley and Gough, Philip},
year = {1990},
month = {01},
pages = {127-160},
title = {The Simple View of Reading},
volume = {2},
journal = {Reading and Writing},
doi = {10.1007/BF00401799}
}

@ARTICLE{ChenCulpepper2020MVP,
  author = {{Chen}, Yinyin and {Culpepper}, Steven A.},
  title = {A Multivariate Probit Model for Learning Trajectories: A Fine-Grained Evaluation of an Educational Intervention},
  journal = {Applied Psychological Measurement},
  year = 2020,
  volume = 44,
  number = {7-8},
  pages = {515--530},
  doi = {10.1177/0146621620920928}
}

@TECHREPORT{ClarkMuthen2009,
  author = {{Clark}, Stephen and {Muth{\'e}n}, Bengt},
  title = {Relating Latent Class Analysis Results to Variables Not Included in the Analysis},
  institution = {Muth{\'e}n \& Muth{\'e}n},
  year = 2009,
  note = {Retrieved from http://www.statmodel.com/download/relatinglca.pdf}
}

@PHDTHESIS{Li2015,
  author = {{Li}, Ming},
  title = "{Investigating Methods of Incorporating Covariates in Growth Mixture Modeling: A Simulation Study}",
  school = {University of Maryland, College Park},
  year = 2015,
  type = {Doctoral dissertation},
  address = {College Park, MD},
  advisor = {{Harring}, Jeffrey~R.}
}

@INCOLLECTION{Croon2002,
  author = {{Croon}, Marcel~A.},
  editor = {{Hagenaars}, Jacques~A. and {McCutcheon}, Allan~L.},
  title = "{Using Predicted Latent Class Membership in Subsequent Analyses}",
  booktitle = {Applied Latent Class Analysis},
  publisher = {Cambridge University Press},
  year = 2002,
  pages = {194--223},
  address = {Cambridge},
  doi = {10.1017/CBO9780511499531.008}
}

@ARTICLE{Bolck2004,
  author = {{Bolck}, Annabel and {Croon}, Marcel and {Hagenaars}, Jacques},
  title = "{Estimating Relationships Between Latent Class Membership and External Variables Using Stepwise Latent Class Analysis}",
  journal = {Political Analysis},
  year = 2004,
  volume = 12,
  number = 1,
  pages = {3--20},
  doi = {10.1093/pan/mgh008}
}

@ARTICLE{Vermunt2010,
  author = {{Vermunt}, Jeroen~K.},
  title = "{Latent Class Modeling with Covariates: Two-Step Estimation Revisited}",
  journal = {Political Analysis},
  year = 2010,
  volume = 18,
  number = 4,
  pages = {450--469},
  doi = {10.1093/pan/mpq014}
}

@ARTICLE{Bakk2013,
  author = {{Bakk}, Zsuzsa and {Tekle}, Fetene~B. and {Vermunt}, Jeroen~K.},
  title = "{Estimating the Association between Latent Class Membership and External Variables Using Bias-Adjusted Three-Step Approaches}",
  journal = {Sociological Methodology},
  year = 2013,
  volume = 43,
  number = 1,
  pages = {272--311},
  doi = {10.1177/0081175012470644}
}

@ARTICLE{Asparouhov2014,
  author = {{Asparouhov}, Tihomir and {Muth{\'e}n}, Bengt},
  title = "{Auxiliary Variables in Latent Class Analysis: Three-Step Approaches}",
  journal = {Structural Equation Modeling: A Multidisciplinary Journal},
  year = 2014,
  volume = 21,
  number = 3,
  pages = {329--341},
  doi = {10.1080/10705511.2014.915181}
}

@ARTICLE{Bakk2018,
  author = {{Bakk}, Zsuzsa and {Kuha}, Jouni},
  title = "{Two-Step Estimation of Models Between Latent Classes and External Variables}",
  journal = {Psychometrika},
  year = 2018,
  volume = 83,
  number = 4,
  pages = {871--892},
  doi = {10.1007/s11336-017-9592-2}
}

@ARTICLE{Ma2025,
        author={Ma, Yawen and Ushakova, Anastasia and Cain, Kate and Wallin, Gabriel},
        title="{A statistical framework for dynamic cognitive diagnosis in digital learning environments}",
        year = 2025,
        month = jun,
        archivePrefix = {arXiv},
        eprint = {2506.14531},
        year = {2025}
}

@ARTICLE{DelaTorre2004,
  author = {{de la Torre}, Jimmy and {Douglas}, Jeffrey~A.},
  title = "{Higher-order Latent Trait Models for Cognitive Diagnosis}",
  journal = {Psychometrika},
  year = 2004,
  volume = 69,
  number = 3,
  pages = {333--353},
  doi = {10.1007/BF02295640}
}

@ARTICLE{Clogg1981,
  author = {{Clogg}, Clifford~C.},
  title = "{New Developments in Latent Structure Analysis}",
  journal = {Evaluation Review},
  year = 1981,
  volume = 5,
  number = 3,
  pages = {445--459},
  doi = {10.1177/0193841X8100500302}
}

@ARTICLE{Dayton1988,
  author = {{Dayton}, C.~Mitchell and {Macready}, George~B.},
  title = "{A Latent Class Covariate Model with Applications to Criterion-Referenced Testing}",
  journal = {Biometrika},
  year = 1988,
  volume = 75,
  number = 1,
  pages = {173--178},
  doi = {10.1093/biomet/75.1.173}
}

@ARTICLE{Lanza2013,
  author = {{Lanza}, Stephanie~T. and {Tan}, Xianming and {Bray}, Beth~C.},
  title = "{Latent Class Analysis with Antenatal Predictors: A One-Step Approach Using Full Information Maximum Likelihood}",
  journal = {Structural Equation Modeling: A Multidisciplinary Journal},
  year = 2013,
  volume = 20,
  number = 4,
  pages = {692--711},
  doi = {10.1080/10705511.2013.824781}
}

@ARTICLE{BandeenRoche1997,
  author = {{Bandeen-Roche}, Karen and {Miglioretti}, Diana~L. and {Zeger}, Scott~L. and {Rathouz}, Paul~J.},
  title = "{Latent Variable Regression for Multiple Discrete Outcomes}",
  journal = {Journal of the American Statistical Association},
  year = 1997,
  month = dec,
  volume = 92,
  number = 440,
  pages = {1375--1386},
  doi = {10.1080/01621459.1997.10473658}
}

@ARTICLE{Xue2002,
  author = {{Xue}, Qian-Li and {Bandeen-Roche}, Karen},
  title = "{Combining Multiple Outcomes to Derive Score-Based Stages of Physical Frailty: The Women’s Health and Aging Study}",
  journal = {Epidemiology},
  year = 2002,
  month = nov,
  volume = 13,
  number = 6,
  pages = {661--670},
  doi = {10.1097/00001648-200211000-00009}
}

@ARTICLE{Bartolucci2014,
  author = {{Bartolucci}, Francesco and {Montanari}, Giorgio~E. and {Pandolfi}, Silvia},
  title = "{Latent Markov Models: A Review of a Class of Models for Longitudinal Data with Discrete Latent Variables}",
  journal = {African Stat. J},
  year = 2014,
  volume = 18,
  pages = {43--82}
}

@BOOK{Collins2010,
  author = {{Collins}, Linda~M. and {Lanza}, Stephanie~T.},
  title = "{Latent Class and Latent Transition Analysis: With Applications in the Behavioral, Social, and Health Sciences}",
  publisher = {John Wiley \& Sons},
  year = 2010,
  address = {Hoboken, NJ},
  doi = {10.1002/9780470567333}
}

@ARTICLE{Muthen2004,
  author = {{Muth{\'e}n}, Bengt},
  title = "{Latent Variable Analysis: Growth Mixture Modeling and Related Techniques for Longitudinal Data}",
  journal = {Handbook of Quantitative Methodology for the Social Sciences},
  year = 2004,
  pages = {345--368},
  publisher = {Sage},
  address = {Thousand Oaks, CA}
}

@ARTICLE{Goodman1974,
  author = {{Goodman}, Leo~A.},
  title = "{Exploratory Latent Structure Analysis Using Both Identifiable and Unidentifiable Models}",
  journal = {Biometrika},
  year = 1974,
  volume = 61,
  number = 2,
  pages = {215--231},
  doi = {10.1093/biomet/61.2.215}
}

@ARTICLE{Nylund2007,
  author = {{Nylund}, Karen~L. and {Asparouhov}, Tihomir and {Muth{\'e}n}, Bengt},
  title = "{Deciding on the Number of Classes in Latent Class Analysis and Growth Mixture Modeling: A Monte Carlo Simulation Study}",
  journal = {Structural Equation Modeling: A Multidisciplinary Journal},
  year = 2007,
  volume = 14,
  number = 4,
  pages = {535--569},
  doi = {10.1080/10705510701575396}
}

@ARTICLE{Wang2018b,
  author  = {Wang, Shiyu and Zhang, Susu and Douglas, Jeff and Culpepper, Steven},
  title   = {Using response times to assess learning progress: A joint model for responses and response times},
  journal = {Measurement: Interdisciplinary Research and Perspectives},
  year    = 2018,
  volume  = 16,
  number  = 1,
  pages   = {45--58},
  doi     = {10.1080/15366367.2018.1435105},
  url     = {https://doi.org/10.1080/15366367.2018.1435105}
}

@ARTICLE{DeLaTorre2011,
  author  = {{de la Torre}, Jimmy},
  title   = "{The generalized DINA model framework}",
  journal = {Psychometrika},
  year    = 2011,
  month   = apr,
  volume  = 76,
  number  = 2,
  pages   = {179--199},
  doi     = {10.1007/s11336-011-9207-7},
  adsurl  = {https://doi.org/10.1007/s11336-011-9207-7}
}

@ARTICLE{Zheng2021,
  author = {Zheng, X. and Yang, J. S. and Harring, J. R.},
  title = {Latent growth curve analysis with item response data: A methodological investigation of model parameterization, estimation, and attrition},
  journal = {Structural Equation Modeling: A Multidisciplinary Journal},
  year = {2021},
  volume = {29},
  number = {2},
  pages = {182--206},
  doi = {10.1080/10705511.2021.1930543},
  url = {https://doi.org/10.1080/10705511.2021.1930543}
}

@ARTICLE{Chen2017,
  author = {{Chen}, Yinghan and {Culpepper}, Steven~Andrew and {Douglas}, Jeffrey~A.},
  title = "{A Hidden Markov Model for Learning Trajectories in Cognitive Diagnosis With Application to Spatial Rotation Skills}",
  journal = {Applied Psychological Measurement},
  year = 2017,
  month = sep,
  volume = 42,
  number = 1,
  pages = {3--19},
  doi = {10.1177/0146621617721250},
  adsurl = {https://doi.org/10.1177/0146621617721250}
}

@ARTICLE{Wang2018a,
  author = {{Wang}, Shiyu and {Yang}, Yan and {Culpepper}, Steven~Andrew and {Douglas}, Jeffrey~A.},
  title = "{Tracking Skill Acquisition With Cognitive Diagnosis Models: A Higher-Order, Hidden Markov Model With Covariates}",
  journal = {Journal of Educational and Behavioral Statistics},
  year = 2018,
  month = jan,
  volume = 43,
  number = 1,
  pages = {57--87},
  doi = {10.3102/1076998617719277},
  adsurl = {https://doi.org/10.3102/1076998617719277}
}

@ARTICLE{Wang2020,
  author = {{Wang}, Daxun and {Cai}, Yan and {Tu}, Dongbo},
  title = "{Q-Matrix Estimation Methods for Cognitive Diagnosis Models: Based on Partial Known Q-Matrix}",
  journal = {Multivariate Behavioral Research},
  year = 2020,
  month = apr,
  volume = 55,
  number = 1,
  pages = {1--13},
  doi = {10.1080/00273171.2020.1746901},
  adsurl = {https://doi.org/10.1080/00273171.2020.1746901}
}

@INPROCEEDINGS{Liang2024ThreeStepLTCDM,
  author = {{Liang}, Q. and {de la Torre}, J. and {Larimer}, M.~E. and {Mun}, E.-Y.},
  title  = "{Mental Health Symptom Profiles Over Time: A Three-Step Latent Transition Cognitive Diagnosis Modeling Analysis with Covariates}",
  booktitle = "{Dependent Data in Social Sciences Research: Forms, Issues, and Methods of Analysis}",
  editor = {{Stemmler}, M. and {Wiedermann}, W. and {Huang}, F.},
  year   = 2024,
  publisher = {Springer},
  address = {New York},
  edition = {2nd},
  pages  = {XXX--XXX},
  doi    = {10.1007/978-3-031-56318-8_22}
}

@ARTICLE{dela2004,
   author = {{de la Torre}, Jimmy and {Douglas}, Jeffrey A.},
    title = "{Higher-Order Latent Trait Models for Cognitive Diagnosis}",
  journal = {Psychometrika},
     year = 2004,
    month = jun,
   volume = 69,
   number = 3,
    pages = {333--353},
      doi = {10.1007/BF02295640},
}

@ARTICLE{Park2014,
   author = {{Park}, Yoon Soo and {Lee}, Young-Sun},
    title = "{An Extension of the DINA Model Using Covariates: Examining Factors Affecting Response Probability and Latent Classification}",
  journal = {Applied Psychological Measurement},
     year = 2014,
    month = oct,
   volume = 38,
   number = 5,
    pages = {556--570},
      doi = {10.1177/0146621614523830},
}

@ARTICLE{Ayers2013,
   author = {{Ayers}, E. and {Rabe-Hesketh}, S. and {Nugent}, R.},
    title = "{Incorporating Student Covariates in Cognitive Diagnosis Models}",
  journal = {Journal of Classification},
     year = 2013,
    month = aug,
   volume = 30,
   number = 2,
    pages = {195--224},
      doi = {10.1007/s00357-013-9130-y},
}

@ARTICLE{templin2006measurement,
  author = {{Templin}, Jonathan~L. and {Henson}, Robert~A.},
   title = "{Measurement of Psychological Disorders Using Cognitive Diagnosis Models}",
 journal = {Psychological Methods},
     year = 2006,
   volume = 11,
   number = 3,
    pages = {287--305},
      doi = {10.1037/1082-989X.11.3.287}
}

@ARTICLE{de2016general,
  author = {{de~la~Torre}, Jimmy and {Chiu}, Chia-Yi},
   title = "{A General Method of Empirical Q-matrix Validation}",
 journal = {Psychometrika},
     year = 2016,
   volume = 81,
   number = 2,
    pages = {253--273},
      doi = {10.1007/s11336-015-9467-8}
}

@manual{MaDeLaTorre2020GDINA,
  title        = {GDINA: An R Package for Cognitive Diagnosis Modeling},
  author       = {Ma, Wen-Chao and de la Torre, Jimmy},
  year         = {2020},
  note         = {R package version 2.9.3},
  url          = {https://CRAN.R-project.org/package=GDINA}
}

@manual{RCoreTeam2023,
  title        = {R: A Language and Environment for Statistical Computing},
  author       = {{R Core Team}},
  organization = {R Foundation for Statistical Computing},
  address      = {Vienna, Austria},
  year         = {2023},
  url          = {https://www.R-project.org/}
}

@article{ChiuDouglas2013ProximityCDM,
  title   = {A nonparametric approach to cognitive diagnosis by proximity to ideal response patterns},
  author  = {Chiu, Chia-Yi and Douglas, Jeffrey A.},
  journal = {Journal of Classification},
  volume  = {30},
  number  = {2},
  pages   = {225--250},
  year    = {2013},
  doi     = {10.1007/s00357-013-9132-9}
}

@article{ZhanJiaoLiao2018CDMRT,
  title   = {Cognitive diagnosis modelling incorporating item response times},
  author  = {Zhan, Peiying and Jiao, Hong and Liao, Dandan},
  journal = {British Journal of Mathematical and Statistical Psychology},
  volume  = {71},
  number  = {2},
  pages   = {262--286},
  year    = {2018},
  doi     = {10.1111/bmsp.12114}
}

@ARTICLE{ChenCulpepper2017DINAQ,
  author  = {Chen, Ying and Culpepper, Steven A.},
  title   = {Bayesian estimation of the {DINA} {Q}-matrix},
  journal = {British Journal of Mathematical and Statistical Psychology},
  year    = {2017},
  volume  = {70},
  number  = {3},
  pages   = {357--380},
  doi     = {10.1111/bmsp.12093}
}

@ARTICLE{Chung2019GibbsQ,
  author  = {Chung, Mengta},
  title   = {A {G}ibbs sampling algorithm that estimates the {Q}-matrix for the {DINA} model},
  journal = {Journal of Mathematical Psychology},
  year    = {2019},
  volume  = {93},
  pages   = {102275},
  doi     = {10.1016/j.jmp.2019.07.002}
}

@ARTICLE{gelman1992inference,
  author = {{Gelman}, Andrew and {Rubin}, Donald B.},
  title = "{Inference from iterative simulation using multiple sequences}",
  journal = {Statistical Science},
  year = 1992,
  volume = {7},
  number = {4},
  pages = {457--472},
  doi = {10.1214/ss/1177011136},
  adsurl = {https://projecteuclid.org/euclid.ss/1177011136}
}

@ARTICLE{MaIaconanglo2016,
       author = {{Ma}, Wenchao and {Iaconangelo}, Charles and {de la Torre}, Jimmy},
        title = "{Model Similarity, Model Selection, and Attribute Classification}",
      journal = {Applied Psychological Measurement},
         year = 2016,
        month = jun,
       volume = {40},
        number = {3},
        pages = {401--416},
          doi = {10.1177/0146621615621717},
       url = {https://doi.org/10.1177/0146621615621717}
}

@article{GuXu2021Identifiability,
  author  = {Gu, Yuqi and Xu, Gongjun},
  title   = {Sufficient and Necessary Conditions for the Identifiability of the Q-Matrix},
  journal = {Statistica Sinica},
  year    = {2021},
  volume  = {31},
  number  = {1},
  pages   = {449--472}
}

@INPROCEEDINGS{Iaconangelo2016ThreeStepCDM,
  author    = {{Iaconangelo}, C. and {de la Torre}, J.},
  title     = "{Three-step estimation of cognitive diagnosis models with covariates}",
  booktitle = {Proceedings of the 81st International Meeting of the Psychometric Society},
  year      = 2016,
  month     = jul,
  address   = {Asheville, NC, United States},
  publisher = {The Psychometric Society},
  pages     = {32},
  note      = {Paper presentation},
  url       = {http://hdl.handle.net/10722/247992}
}

@ARTICLE{Zhan2019JEBStats,
  author = {{Zhan}, Peida and {Jiao}, Hong and {Man}, Kaiwen},
  title = "{Using JAGS for Bayesian Cognitive Diagnosis Modeling: A Tutorial}",
  journal = {Journal of Educational and Behavioral Statistics},
  archivePrefix = "arXiv",
  eprint = {1708.02632},
  primaryClass = "stat.CO",
  year = 2019,
  volume = 44,
  number = 4,
  pages = {473--503},
  doi = {10.3102/1076998619826040},
}

@ARTICLE{LiHarring2017,
  author = {{Li}, Ming and {Harring}, Jeffrey~R.},
  title = "{Investigating Approaches to Estimating Covariate Effects in Growth Mixture Modeling: A Simulation Study}",
  journal = {Educational and Psychological Measurement},
  year = 2017,
  volume = 77,
  number = 5,
  pages = {766--791},
  doi = {10.1177/0013164416653789}
}

@ARTICLE{DiMari2016ThreeStepLM,
  author = {{Di Mari}, Roberto and {Oberski}, Daniel L. and {Vermunt}, Jeroen K.},
  title = "{Bias-Adjusted Three-Step Latent Markov Modeling With Covariates}",
  journal = {Structural Equation Modeling: A Multidisciplinary Journal},
  year = 2016,
  volume = 23,
  number = 5,
  pages = {649--666},
  doi = {10.1080/10705511.2016.1191015},
  url = {https://doi.org/10.1080/10705511.2016.1191015}
}

@ARTICLE{YigitDouglas2021FirstOrder,
   author = {{Yigit}, Hulya D. and {Douglas}, Jeffrey A.},
   title = "{First-Order Learning Models With the G-DINA: Estimation With the EM Algorithm and Applications}",
   journal = {Journal of Educational and Behavioral Statistics},
     year = 2021,
   volume = 45,
   number = 3,
    pages = {264--302},
      doi = {10.1177/0146621621990746}
}

@ARTICLE{Liang2023ThreeStep,
   author = {{Liang}, Qianru and {de la Torre}, Jimmy and {Law}, Nancy},
   title = "{Latent Transition Cognitive Diagnosis Model With Covariates: A Three-Step Approach}",
   journal = {Journal of Educational and Behavioral Statistics},
   year = 2023,
   volume = 48,
   number = 6,
   pages = {690--718},
   doi = {10.3102/10769986231163320},
}

\appendix

\section{Code Availability}
\label{sec:app_code}

The implementation of the joint estimation framework using R (version 4.5.2; \cite{RCoreTeam2023}) is available at \url{https://github.com/Yawen-Ma/IMPS_proceeding2025}. The stepwise procedure builds on the R implementation of \citep{Liang2023ThreeStep}. 
A minor modification was made to ensure proper normalisation of the transition probability matrix. 
The corrected code is also available at the author's GitHub. 

\section{Simulation Evaluation Metrics}
\label{sec:metric}

Model performance was evaluated with respect to both parameter estimation accuracy and classification accuracy. 
For item parameters and regression coefficients in the initial attribute and transition models, estimation accuracy was quantified using the mean absolute error (MAE) and root mean square error (RMSE) across replications, defined as
$$
\text{MAE} = \frac{1}{R} \sum_{r=1}^{R} \left| \hat{p}^{(r)} - p \right|, \quad
\text{RMSE} = \sqrt{ \frac{1}{R} \sum_{r=1}^{R} \left( \hat{p}^{(r)} - p \right)^2 },
$$
where $p$ denotes the true parameter value, $\hat{p}^{(r)}$ is the estimate obtained in replication $r$, and $R$ is the total number of replications.

Classification accuracy was evaluated at the attribute level using the attribute accuracy rate (AAR). 
For each attribute $k$ at a given time point, AAR was computed as
$$
\text{AAR}_{k,t} = \frac{1}{NR} \sum_{r=1}^{R} \sum_{i=1}^{N} \mathbb{I} \bigl(\hat{\alpha}^{(r)}_{ikt} = \alpha_{ikt}\bigr),
$$
where $N$ is the number of examinees, $\alpha_{ik}$ and $\hat{\alpha}_{ik}$ denote the true and estimated mastery status of attribute $k$ for examinee $i$, respectively, and $\mathbb{I}(\cdot)$ is an indicator function equal to 1 when the condition holds and 0 otherwise.

\section{Item Parameter Recovery}
\label{sec:app_item}

\begin{table}[!h]
\centering
\caption{Estimation accuracy of item parameters ($g,s$) evaluated by MAE and RMSE across sample size ($N$) and test length ($J$).}
\label{tab:gs_bias}
 {
\begin{tabular}{ccc|cccc|cccc}
\toprule
\multicolumn{3}{c}{} & \multicolumn{4}{c}{Joint} & \multicolumn{4}{c}{Stepwise} \\
\cmidrule(l{3pt}r{3pt}){4-7} \cmidrule(l{3pt}r{3pt}){8-11}
$N$ & $J$ & Metric &
$g_{t1}$ & $g_{t2}$ & $s_{t1}$ & $s_{t2}$ &
$g_{t1}$ & $g_{t2}$ & $s_{t1}$ & $s_{t2}$ \\
\midrule
200 & 6  & MAE
& 0.060 & 0.048 & 0.042 & 0.058
& 0.055 & 0.062 & 0.049 & 0.059 \\
     &    & RMSE
& 0.073 & 0.061 & 0.053 & 0.069
& 0.069 & 0.077 & 0.061 & 0.072 \\
\addlinespace
400 & 18 & MAE
& 0.026 & 0.022 & 0.022 & 0.027
& 0.023 & 0.023 & 0.023 & 0.024 \\
     &    & RMSE
& 0.032 & 0.028 & 0.027 & 0.034
& 0.028 & 0.030 & 0.028 & 0.030 \\
\addlinespace
600 & 30 & MAE
& 0.019 & 0.022 & 0.018 & 0.019
& 0.020 & 0.022 & 0.018 & 0.018 \\
     &    & RMSE
& 0.024 & 0.027 & 0.022 & 0.023
& 0.024 & 0.027 & 0.022 & 0.023 \\
\bottomrule
\end{tabular}}
\end{table}

\section{Simulation Q-Matrices}
\label{sec:app_Qmatrix}
The true Q-matrices used in the simulation study for 
$(N, J) = (200, 6)$, $(400, 18)$, and $(600, 30)$ are provided here.

\begin{table}[!ht]
\centering
\caption{True Q-matrices with $K=2$ attributes under different test lengths: (a) $J=6$, (b) $J=18$, and (c) $J=30$.}
\label{tab:true_qmatrix_all}

\begin{tabular}{ccc}
\begin{tabular}{c|cc}
\toprule
Item & $A_1$ & $A_2$ \\
\midrule
1 & 1 & 0 \\
2 & 0 & 1 \\
3 & 1 & 0 \\
4 & 0 & 1 \\
5 & 1 & 0 \\
6 & 1 & 1 \\
\bottomrule
\end{tabular}
&
\begin{tabular}{c|cc}
\toprule
Item & $A_1$ & $A_2$ \\
\midrule
1  & 1 & 0 \\
2  & 0 & 1 \\
3  & 1 & 0 \\
4  & 0 & 1 \\
5  & 1 & 0 \\
6  & 0 & 1 \\
7  & 1 & 0 \\
8  & 0 & 1 \\
9  & 1 & 0 \\
10 & 0 & 1 \\
11 & 1 & 0 \\
12 & 0 & 1 \\
13 & 1 & 0 \\
14 & 0 & 1 \\
15 & 1 & 1 \\
16 & 1 & 1 \\
17 & 1 & 1 \\
18 & 1 & 1 \\
\bottomrule
\end{tabular}
&
\begin{tabular}{c|cc}
\toprule
Item & $A_1$ & $A_2$ \\
\midrule
1  & 1 & 0 \\
2  & 0 & 1 \\
3  & 1 & 0 \\
4  & 0 & 1 \\
5  & 1 & 0 \\
6  & 0 & 1 \\
7  & 1 & 0 \\
8  & 0 & 1 \\
9  & 1 & 0 \\
10 & 0 & 1 \\
11 & 1 & 0 \\
12 & 0 & 1 \\
13 & 1 & 0 \\
14 & 0 & 1 \\
15 & 1 & 0 \\
16 & 0 & 1 \\
17 & 1 & 0 \\
18 & 0 & 1 \\
19 & 1 & 0 \\
20 & 0 & 1 \\
21 & 1 & 0 \\
22 & 0 & 1 \\
23 & 1 & 0 \\
24 & 0 & 1 \\
25 & 1 & 1 \\
26 & 1 & 1 \\
27 & 1 & 1 \\
28 & 1 & 1 \\
29 & 1 & 1 \\
30 & 1 & 1 \\
\bottomrule
\end{tabular}
\end{tabular}

\vspace{0.5em}

\begin{tabular}{ccc}
\small (a) $J=6$ & \small (b) $J=18$ & \small (c) $J=30$
\end{tabular}

\end{table}

\section{Sensitivity Analysis with Varying Levels of Covariate Correlation}
\label{sec:sensitivity_cov}
 {
Sensitivity analyses for the most challenging simulation condition, $(N, J) = (200, 6)$, under varying correlation levels $\rho \in \{0, 0.2, 0.4, 0.6, 0.8\}$ are reported in Tables \ref{tab:cov} and \ref{tab:cov1}. Panels A-D summarise the results for attribute profiles, item parameters, initial attribute parameters, and transition parameters, respectively. The case $\rho = 0$ corresponds to the independent covariate setting, where the covariate vector $\mathbf{Z}_i$ follows a multivariate normal distribution, $\mathbf{Z}_i \sim \mathcal{N}(\mathbf{0}, \mathbf{I}_3)$. Across $\rho \in {0,0.2,0.4,0.6,0.8}$, the differences in AAR were generally small, typically within about 0.01–0.02 (Table \ref{tab:cov}, Panel A). A similar pattern was observed for item parameter recovery (Table \ref{tab:cov}, Panel B), where the estimation accuracy of the guessing and slipping parameters remained stable across correlation levels, with differences in MAE and RMSE typically within approximately 0.01–0.02. }

 {As shown in Panels C and D of Table \ref{tab:cov1}, the estimation of covariate effects exhibits a similar pattern across correlation levels. For the initial covariate effects (Panel C), under the joint approach, the variation in MAE and RMSE across $\rho$ values is relatively small, typically within about 0.03–0.10 for intercept parameters and less than 0.03 for covariate effects. A similar pattern is observed for the stepwise method, where the variation across $\rho$ is also small, generally within about 0.05–0.10 for intercept parameters and less than 0.05 for covariate effects. For the transition covariate effects (Panel D), the stepwise method exhibits substantially larger estimation bias overall. In addition, as $\rho$ varies, the corresponding changes in MAE and RMSE are approximately 0.2–0.6.}

\begin{table}[!htbp]
\centering
\caption{ {Sensitivity analysis results under varying covariate correlation levels.}}
\label{tab:cov}
\begin{threeparttable}
\small

\textbf{ {Panel A. Attribute recovery (AAR)}}\\[3pt]
 {
\begin{tabular}{cccc|cc|cc}
\toprule
\multicolumn{4}{c}{} & \multicolumn{2}{c}{Joint} & \multicolumn{2}{c}{Stepwise} \\
\cmidrule(l{3pt}r{3pt}){5-6} \cmidrule(l{3pt}r{3pt}){7-8}
$\rho$ & $N$ & $J$ & $T$ & AAR$_1$ & AAR$_2$ & AAR$_1$ & AAR$_2$ \\
\midrule
0 & 200 & 6 & 1 & 0.890 & 0.860 & 0.863 & 0.842 \\
    &  &  & 2 & 0.898 & 0.854 & 0.859 & 0.847 \\
0.2 & 200 & 6 & 1 & 0.901 & 0.863 & 0.873 & 0.843 \\
    &  &  & 2 & 0.894 & 0.849 & 0.879 & 0.853 \\
0.4 & 200 & 6 & 1 & 0.899 & 0.869 & 0.868 & 0.846 \\
    &  &  & 2 & 0.891 & 0.858 & 0.861 & 0.843 \\
0.6 & 200 & 6 & 1 & 0.904 & 0.866 & 0.870 & 0.835 \\
    &  &  & 2 & 0.890 & 0.843 & 0.863 & 0.849 \\
0.8 & 200 & 6 & 1 & 0.898 & 0.877 & 0.874 & 0.847 \\
    &     &  & 2 & 0.885 & 0.855 & 0.853 & 0.845 \\
\bottomrule
\end{tabular}}

\vspace{1em}

\textbf{ {Panel B. Item parameter recovery}}\\[3pt]
 {
\begin{tabular}{cccc|cccc|cccc}
\toprule
\multicolumn{4}{c}{} & \multicolumn{4}{c}{Joint} & \multicolumn{4}{c}{Stepwise} \\
\cmidrule(l{3pt}r{3pt}){5-8} \cmidrule(l{3pt}r{3pt}){9-12}
$\rho$ & $N$ & $J$ & Metric &
$g_{t1}$ & $g_{t2}$ & $s_{t1}$ & $s_{t2}$ &
$g_{t1}$ & $g_{t2}$ & $s_{t1}$ & $s_{t2}$ \\
\midrule
0 & 200 & 6 & MAE
& 0.064 & 0.047 & 0.043 & 0.059
& 0.066 & 0.055 & 0.058 & 0.059 \\
    &     &   & RMSE
& 0.073 & 0.058 & 0.051 & 0.070
& 0.080 & 0.067 & 0.068 & 0.068 \\
\addlinespace
0.2 & 200 & 6 & MAE
& 0.049 & 0.047 & 0.041 & 0.068
& 0.062 & 0.053 & 0.055 & 0.045 \\
    &     &   & RMSE
& 0.059 & 0.056 & 0.053 & 0.081
& 0.076 & 0.064 & 0.066 & 0.052 \\
\addlinespace
0.4 & 200 & 6 & MAE
& 0.060 & 0.048 & 0.042 & 0.058
& 0.055 & 0.062 & 0.049 & 0.059 \\
    &     &   & RMSE
& 0.073 & 0.061 & 0.053 & 0.069
& 0.069 & 0.077 & 0.061 & 0.072 \\
\addlinespace
0.6 & 200 & 6 & MAE
& 0.048 & 0.044 & 0.039 & 0.071
& 0.067 & 0.050 & 0.053 & 0.048 \\
    &     &   & RMSE
& 0.057 & 0.050 & 0.047 & 0.084
& 0.081 & 0.063 & 0.062 & 0.057 \\
\addlinespace
0.8 & 200 & 6 & MAE
& 0.065 & 0.050 & 0.041 & 0.061
& 0.062 & 0.058 & 0.058 & 0.056 \\
    &     &   & RMSE
& 0.077 & 0.062 & 0.050 & 0.070
& 0.077 & 0.073 & 0.067 & 0.069 \\
\bottomrule
\end{tabular}}
\end{threeparttable}
\end{table}

\begin{table}[!htbp]
\centering
\caption{ {Sensitivity analysis results under varying covariate correlation levels for regression parameters.}}
\label{tab:cov1}
\begin{threeparttable}
\small
\textbf{ {Panel C. Initial covariate effects}}\\[3pt]
 {
\begin{tabular}{cccc|cccc|cccc}
\toprule
\multicolumn{4}{c}{} & \multicolumn{4}{c}{Joint} & \multicolumn{4}{c}{Stepwise} \\
\cmidrule(l{3pt}r{3pt}){5-8} \cmidrule(l{3pt}r{3pt}){9-12}
$\rho$ & $N$ & $J$ & Metric
& $\beta_{0,1}$ & $\beta_{0,2}$ & $\beta_{Z,1}$ & $\beta_{Z,2}$
& $\beta_{0,1}$ & $\beta_{0,2}$ & $\beta_{Z,1}$ & $\beta_{Z,2}$ \\
\midrule
0 & 200 & 6 & MAE
& 0.206 & 0.304 & 0.097 & 0.071
& 0.300 & 0.279 & 0.115 & 0.082 \\
    &     &   & RMSE
& 0.231 & 0.371 & 0.113 & 0.087
& 0.383 & 0.364 & 0.140 & 0.109 \\
\addlinespace
0.2 & 200 & 6 & MAE
& 0.080 & 0.362 & 0.057 & 0.073
& 0.226 & 0.251 & 0.054 & 0.065 \\
    &     &   & RMSE
& 0.115 & 0.430 & 0.073 & 0.083
& 0.256 & 0.380 & 0.073 & 0.081 \\
\addlinespace
0.4 & 200 & 6 & MAE
& 0.182 & 0.366 & 0.055 & 0.095
& 0.229 & 0.279 & 0.060 & 0.105 \\
    &     &   & RMSE
& 0.197 & 0.425 & 0.067 & 0.105
& 0.275 & 0.317 & 0.067 & 0.119 \\
\addlinespace
0.6 & 200 & 6 & MAE
& 0.097 & 0.301 & 0.059 & 0.073
& 0.214 & 0.303 & 0.080 & 0.080 \\
    &     &   & RMSE
& 0.127 & 0.359 & 0.069 & 0.082
& 0.262 & 0.442 & 0.092 & 0.091 \\
\addlinespace
0.8 & 200 & 6 & MAE
& 0.160 & 0.301 & 0.057 & 0.053
& 0.247 & 0.225 & 0.051 & 0.050 \\
    &     &   & RMSE
& 0.184 & 0.347 & 0.064 & 0.067
& 0.255 & 0.277 & 0.055 & 0.064 \\
\bottomrule
\end{tabular}}

\vspace{1em}
\textbf{ {Panel D. Transition covariate effects}}\\[3pt]
 {
\begin{tabular}{cccc|cccc|cccc}
\toprule
\multicolumn{4}{c}{} & \multicolumn{4}{c}{Joint} & \multicolumn{4}{c}{Stepwise} \\
\cmidrule(l{3pt}r{3pt}){5-8} \cmidrule(l{3pt}r{3pt}){9-12}
$\rho$ & $N$ & $J$ & Metric
& $\gamma_{01,0,1}$ & $\gamma_{01,0,2}$ & $\gamma_{01,Z,1}$ & $\gamma_{01,Z,2}$
& $\gamma_{01,0,1}$ & $\gamma_{01,0,2}$ & $\gamma_{01,Z,1}$ & $\gamma_{01,Z,2}$ \\
\midrule
0 & 200 & 6 & MAE
& 0.085 & 0.175 & 0.105 & 0.111
& 1.254 & 1.548 & 0.835 & 0.746 \\
    &     &   & RMSE
& 0.102 & 0.211 & 0.114 & 0.127
& 1.387 & 1.793 & 1.147 & 0.817 \\
\addlinespace
0.2 & 200 & 6 & MAE
& 0.120 & 0.177 & 0.101 & 0.086
& 1.716 & 1.598 & 0.944 & 0.732 \\
    &     &   & RMSE
& 0.143 & 0.199 & 0.125 & 0.110
& 2.032 & 1.655 & 1.100 & 0.968 \\
\addlinespace
0.4 & 200 & 6 & MAE
& 0.122 & 0.210 & 0.100 & 0.140
& 1.263 & 1.468 & 1.008 & 0.805 \\
    &     &   & RMSE
& 0.146 & 0.246 & 0.125 & 0.164
& 1.352 & 1.542 & 1.208 & 0.911 \\
\addlinespace
0.6 & 200 & 6 & MAE
& 0.103 & 0.192 & 0.114 & 0.079
& 1.639 & 1.427 & 0.672 & 0.535 \\
    &     &   & RMSE
& 0.122 & 0.204 & 0.126 & 0.104
& 1.931 & 1.542 & 0.791 & 0.745 \\
\addlinespace
0.8 & 200 & 6 & MAE
& 0.109 & 0.228 & 0.092 & 0.135
& 1.209 & 1.523 & 0.704 & 0.623 \\
    &     &   & RMSE
& 0.137 & 0.262 & 0.106 & 0.152
& 1.280 & 1.890 & 0.831 & 0.776 \\
\bottomrule
\end{tabular}}

\end{threeparttable}
\end{table}

\end{document}